\documentclass[letterpaper,11pt]{article}
\usepackage{amssymb}

\catcode`\@=11
\def\newexample#1{\@ifnextchar[{\@oexm{#1}}{\@nexm{#1}}}

\def\@nexm#1#2{%
\@ifnextchar[{\@xnexm{#1}{#2}}{\@ynexm{#1}{#2}}}

\def\@xnexm#1#2[#3]{\expandafter\@ifdefinable\csname #1\endcsname
{\@definecounter{#1}\@addtoreset{#1}{#3}%
\expandafter\xdef\csname the#1\endcsname{\expandafter\noexpand
  \csname the#3\endcsname \@exmcountersep \@exmcounter{#1}}%
\global\@namedef{#1}{\@exm{#1}{#2}}\global\@namedef{end#1}{\@endexample}}}

\def\@ynexm#1#2{\expandafter\@ifdefinable\csname #1\endcsname
{\@definecounter{#1}%
\expandafter\xdef\csname the#1\endcsname{\@exmcounter{#1}}%
\global\@namedef{#1}{\@exm{#1}{#2}}\global\@namedef{end#1}{\@endexample}}}

\def\@oexm#1[#2]#3{\expandafter\@ifdefinable\csname #1\endcsname
  {\global\@namedef{the#1}{\@nameuse{the#2}}%
\global\@namedef{#1}{\@exm{#2}{#3}}%
\global\@namedef{end#1}{\@endexample}}}

\def\@exm#1#2{\refstepcounter
    {#1}\@ifnextchar[{\@yexm{#1}{#2}}{\@xexm{#1}{#2}}}

\def\@xexm#1#2{\@beginexample{#2}{\csname the#1\endcsname}\ignorespaces}
\def\@yexm#1#2[#3]{\@opargbeginexample{#2}{\csname
       the#1\endcsname}{#3}\ignorespaces}

\def\@exmcounter#1{\noexpand\arabic{#1}}
\def\@exmcountersep{.}
\def\@beginexample#1#2{\trivlist \item[\hskip 
\labelsep{\bf #1\ #2:}]}
\def\@opargbeginexample#1#2#3{\trivlist
      \item[\hskip \labelsep{\bf #1\ #2\ }#3{\bf :}]}
\def\@endexample{\endtrivlist}

\catcode`\@=12

\newtheorem{thm}{{\bf Theorem}}[section]
\newtheorem{df}{{\bf Definition}}[section]

\newtheorem{rem}{{\bf Remark}}[section]

\newtheorem{claim}{{\bf Claim}}[section]

\newcommand{\leng}[1]{|{#1}|}
\newcommand{\BQED}{\hfill \hbox{\rule{8pt}{8pt}}}
\newcommand{\QED}{\hfill \mbox{$\square$}}

\newcommand{\msc}[1]{\mbox{{\sc #1}}}
\newcommand{\bm}[1]{\mbox{\boldmath{$#1$}}}

\newenvironment{namelist}[1]{%
\begin{list}{}
  { 
	\settowidth{\labelwidth}{#1}
	\setlength{\leftmargin}{1.1\labelwidth}}
  \setlength{\itemsep}{0cm}
}{%
\end{list}}

\catcode`\@=11
\renewcommand{\@biblabel}[1]{\hspace*{\fill}[#1]}
\catcode`\@=12
\begin{document}
\begin{center}
{\Large {\bf Approximation Algorithms for the Highway Problem}}\smallskip\\
{\Large {\bf under the Coupon Model}\footnote[1]{~This research was 
supported in part by JSPS Global COE program ``Computationism as a 
Foundation for the Sciences.''}}\medskip\\
\begin{tabular}{ccc}
{\sc Ryoso Hamane} & {\sc Toshiya Itoh} & {\sc Kouhei Tomita}\\
{\sf hamane@dac.gsic.titech.ac.jp} & 
{\sf titoh@dac.gsic.titech.ac.jp} & {\sf tomita@dac.gsic.titech.ac.jp}\medskip\\
\multicolumn{3}{c}{{\sf Tokyo Institute of Technology}}
\end{tabular}
\end{center}\medskip
{\bf Abstract:} When a store sells items to customers, 
the store wishes to decide the prices~of~items~to maximize 
its profit. Intuitively, if the store sells the items with low 
(resp. high) prices, the customers~buy~more~(resp. less) items, 
which provides less profit to the store. So it would be hard 
for the store to decide the prices~of items. 
Assume that the store has a set $V$ of $n$ items and 
there~is~a~set~$E$~of~$m$~customers~who~wish~to~buy~the  
items, and also assume that 
each item $i \in V$ has the production cost $d_{i}$~and~each~customer 
$e_{j} \in E$~has~the valuation $v_{j}$ on the bundle 
$e_{j} \subseteq V$ of items. 
When~the~store sells an item $i \in V$~at~the~price $r_{i}$,~the~profit~for 
the item~$i$~is~$p_{i}=r_{i}-d_{i}$.~The~goal~of~the~store~is~to~decide 
the price~of~each~item~to~maximize~its~total~profit. We refer to 
this maximization problem~as~the {\it item pricing\/} problem. 
In most of the previous works,~the~item pricing problem was considered 
under the assumption that $p_{i} \geq 0$ for each $i \in V$, however, 
Balcan,~et~al.~[In Proc. of WINE, LNCS 4858, 2007] introduced the notion 
of ``loss-leader,'' and showed that the seller can get~more~total~profit 
in the case that $p_{i} < 0$ is allowed than in the case that 
$p_{i} < 0$~is~not~allowed.~In~this~pa\-per, we consider the line highway problem 
(in which each customer is interested in an interval~on~the~line~of the items) 
and the cycle highway 
problem (in which each customer is interested in an interval 
on~the~cycle~of the items), and show approximation 
algorithms for the line highway problem and~the~cycle~highway~prob\-lem 
in which the smallest valuation 
is $s$ and the largest valuation is $\ell$ 
(this is called an $[s,\ell]$-valuation~set\-ting) 
or all valuations are identical (this is called a single 
valuation setting).\medskip
\section{Introduction} \label{intro}
%
\subsection{Background} \label{background}
%
When a store sells items to customers, the store wishes 
to decide the prices of items~to~maximize~its profit. Intuitively, 
if the store sells the items with low (resp. high) prices, then 
the customers~buy~more (resp. less) items, 
which provides less profit to the store. 
So it would be hard for the store~to~decide~the~prices~of~items. 
Assume that the store has a set $I=\{1,2,\ldots,n\}$ of $n$ items and 
there~is~a~set~$C=\{c_{1},c_{2},\ldots,c_{m}\}$~of~$m$~customers who wish to 
buy the items. 
The goal of the store is to decide 
the price of each item to maximize~its profit. We refer to 
this problem as the {\it item~pricing\/} problem. 
We classify the item pricing~problem~accord\-ing to how 
many items the store can sell and how the customers 
valuate items.~If~the~store~can~sell~each~item $i$ 
with unlimited (resp. limited) amount, we refer to this as 
the {\it unlimited\/} (resp. {\it limited\/})~supply~model.~The 
item pricing problem is said~to~be~{\it single-minded\/} \cite{Getal} 
if each customer $c_{j} \in C$ is interested 
in~only~a~single~bundle~$e_{j} = \{j_{1},j_{2},\ldots,\} \subseteq I$~of~items~with 
valuation $v_{j}\geq 0$ and has valuation~``0''~on~all~other~bundles of items. 
We say that the item pricing problem is 
{\it unit-demand\/} \cite{Getal} if each customer $c_{j} \in C$ assigns 
valua\-tion $v_{j}^{i}\geq 0$ to each item $i \in I$ and 
buys one of~the most beneficial items for $c_{j} \in C$. 

By regarding the set $I$ of $n$ items as the set $V$
of $n$ vertices and the set $C$ of $m$ customers~as~the set $E$ of $m$ hyperedges, 
each of which has weight $v_{j}\geq 0$, 
this can be formulated~by~a~{\it weighted\/}~hypergraph $\tilde{G}=(V,E,\{v_{j}\})$. 
Note that the hypergraph $\tilde{G}$ might have selfloops (corresponding to 
customers~that are interested in a single item) and multiedges 
(corresponding to customers that want~to~get~the~same~bundle of items). 
For a weighted hypergraph $\tilde{G}=(V,E,\{v_{j}\})$, 
assume~that~each~item~$i \in V$~has~the 
{\it production\/} cost $d_{i}$ and each customer $e_{j} \in E$ has the 
valuation $v_{j}$. For $\tilde{G}$, we define~a~{\it reduced\/}~instance 
$G=(V,E,\{w_{j}\})$ to be 
$w_{j} = v_{j}- \sum_{i \in e_{j}} d_{i}$ for each $e_{j} \in E$. 
If an item $i \in V$ is assigned a price~$p_{i}$~in~the reduced 
instance $G$, then its selling price is given by $r_{i}=p_{i}+d_{i}$. 
In this paper, 
we focus~on~the~single-minded and unlimited supply model and consider 
reduced instances $G$'s~of~weighted~hypergraphs. 
We~say~that~$G=(V,E,\{w_{j}\})$~is~an~in\-stance of the 
{\it $k$-hypergraph vertex pricing\/} problem if 
$\leng{e_{j}} \leq k$ for each $e_{j} \in E$, 
an instance of the {\it graph~vertex pricing\/}~problem 
if $\leng{e_{j}} \leq 2$ for each $e_{j} \in E$,~and~an~instance 
of the {\it bipartite graph vertex~pricing\/}~problem~if 
$G$ is a bipartite graph. 
As~a~special~case~of~the~hypergraph~vertex~pricing~problem, 
we~also~say~that~$G=(V,E,\{w_{j}\})$ is an instance of the 
{\it highway\/} problem if each $e_{j} \in E$ is an interval on $V$ 
(the definition will be given in Definition \ref{df-line-highway} for 
the {\it line\/} highway problem and in Definition 
\ref{df-cycle-highway} for the {\it cycle\/} highway problem). 

In most of the previous works \cite{BB,BK06,BK07,Getal}, the item pricing problem 
is considered under~the~model~that $p_{i} \geq 0$ for each item $i \in V$ 
(this is called the {\it positive price\/} model). 
By~introducing~the~notion~of~{\it loss-leader\/} \cite{D}, however, 
Balcan,~et~al.~\cite{BBCH} consider several price models 
in which $p_{i} < 0$~for~some~item~$i \in V$~(these~are referred to as the 
{\it discount\/} model, the {\it $B$-bounded discount\/} model, 
the {\it coupon\/} model, etc., and are 
formally defined in Subsection \ref{subsec-models}), and showed that 
the seller could 
get more profit in the~case~that~$p_{i} <0$~is~allowed 
than in the case that $p_{i}<0$ is not allowed. 
%
\subsection{Related Works} \label{related}
%
\subsubsection{Positive Price Models} \label{subsec-positive}
%
For the hypergraph vertex pricing problem,~Guruswami,~et~al.~\cite[Theorem~5.2]{Getal} 
show an $O(\log m+\log n)$-approximation algorithm. On the other hand, 
Demaine,~et~al.~\cite[Theorem 3.2]{DFHS} 
present that~it~is~hard~to~approximate the hypergraph vertex pricing problem 
within a factor of $\log^{\delta} n$ for some $\delta>0$~under~the~assumption 
that ${\rm NP} \not \subseteq {\rm BPTIME}(2^{n^{\epsilon}})$ 
for some $\epsilon > 0$. 
For the $k$-hypergraph 
vertex pricing problem,~Briest and Krysta \cite[Theorem 5.1]{BK06} show an 
$O(k^{2})$-approximation algorithm, which is improved~to~an~$O(k)$-ap\-proximation 
algorithm \cite[Theorem~2]{BB}. 
For the graph vertex pricing problem, 
Balcan~and~Blum~derive~a $1/4$-approximation algorithm \cite[Theorem 1]{BB}, 
while by the reduction from the vertex cover, 
Guruswami,~et al. \cite[Theorem~3.1]{Getal} 
show that the graph vertex pricing problem is APX-hard even when 
all valuations~are identical (if selfloops are allowed) or all valuations are 
either 1 or 2 (if selfloops are not allowed). 
For the highway problem, 
Balcan and Blum \cite[Theorem 3]{BB} show an $O(\log n)$-approximation algorithm 
and~for~the highway problem that forms a hierarchy, Balcan and 
Blum \cite[Theorem 4]{BB} show a fully~polynomial~time~ap\-proximation scheme. 
For the nonapproximability 
for the highway problem, see \cite{BK06,GLSU}.
%
\subsubsection{Other Models Based on Loss-Leader} \label{subsec-other}
%
For the highway problem, we know   
the $\Omega(\log n)$ gap between the positive price model 
and~the~($B$-bounded) discount model \cite[Theorem 1]{BBCH-TR}, 
and the $\Omega(\log n)$ gap between the coupon model 
and the ($B$-bounded)~dis\-count model \cite[Theorem 2]{BBCH-TR}. 
For the graph vertex pricing problem, 
the $\Omega(\log n)$ gap between the positive price model and the 
$B$-bounded discount model \cite[Theorem 3]{BBCH-TR} is known. 
For the highway problem, Balcan, et al. \cite[Theorem 3]{BBCH} 
show a 2.33-approximation algorithm 
under~the~coupon~model~if~all~valuations~are~i\-dentical 
and for the highway problem on tree, 
Balcan,~et~al.~\cite[Theorem~15]{BBCH-TR}~show~a~4-approximation~algo\-rithm 
under the coupon model if all valuations are identical. 
%
\subsection{Main Results} \label{main}
%
In this paper, we consider the highway problem with {\it $[s,\ell]$-valuation\/}, 
which is the highway problem~with~the smallest valuation $s$ and the largest 
valuation $\ell$. We also classify the highway problem into the {\it line\/} 
highway problem and the {\it cycle\/} highway problem in which each interval 
is defined on the line~of~items~and~the~cycle of items, respectively. 
Then we consider the line highway problem with 
$[s,\ell]$-valuation~and~the~cycle~highway problem with 
$[s,\ell]$-valuation and a single valuation.\medskip\\
{\bf Theorem \ref{thm-line-highway}:} {\sl 
On an instance $G=(V,E,\{w_{j}\})$ of the line highway problem 
with $[s,\ell]$-valuation,~the~algo\-rithm $\msc{Line}_{[s,\ell]}$ 
outputs a price vector $\bm{p}$ that satisfies
\[
\frac{\msc{Opt}_{\rm coup}(G)}{{\bf E}[\msc{Profit}_{\rm coup}(\bm{p})]} 
\leq \left\{
\begin{array}{cl}
4(1-\ln r) & 0 \leq r \leq \alpha,~1/\sqrt{e} \leq r \leq 1;\\
3/r & \alpha < r \leq 1/2;\\
6 & 1/2 < r < 1/\sqrt{e}, 
\end{array} \right.
\]
where $r=s/\ell$ is the ratio between the smallest and the largest valuations and 
$\alpha \approx 0.3824$, i.e.,~$\alpha$~is~the~solution 
of the equality $3/x =4(1-\ln x)$.}\medskip\\
{\bf Theorem \ref{thm-cycle-[s,l]}:} {\sl 
On an instance $G=(V,E,\{w_{j}\})$ of the cycle highway problem 
with $[s,\ell]$-valuation,~the~algorithm $\msc{Cycle}_{[s,\ell]}$ 
outputs a price vector $\bm{p}$ that satisfies
\[
\msc{Opt}_{\rm coup}(G) \leq 4(1-\ln r) \cdot 
{\bf E}[\msc{Profit}_{\rm coup}(\bm{p})], 
\]
%
%
%
%
%
%
%
where $r=s/\ell$ is the ratio between the smallest and the largest valuations 
and $\alpha \approx 0.3824$,~i.e.,~$\alpha$~is~the~solution of the equality 
$3/x =4(1-\ln x)$.}\medskip\\
{\bf Theorem \ref{thm-cycle-single-val}:} {\sl 
On an instance $G=(V,E)$ of the cycle highway problem 
with a single valuation,~the~algo\-rithm {\sc Cyc\_Single\_Val} 
outputs a price vector $\bm{p}$ that satisfies}
\[
\msc{Opt}_{\rm coup}(G) \leq 2.747 \cdot \msc{Profit}_{\rm coup}(\bm{p}).
\]
For the line highway problem, Theorem \ref{thm-line-highway} is 
an extension of the 2.33-approximation algorithm~with~a~single valuation 
due to Balcan, et al. \cite[Theorem 3]{BBCH}. The cycle highway problem is 
introduced~in~this~paper~as a generalization of the line highway problem, and 
Theorem \ref{thm-cycle-single-val} 
can be regarded as an extension~of~the~2.33-approximation 
algorithm for the line highway problem with a single valuation 
\cite[Theorem 3]{BBCH}. 
%
\section{Preliminaries} \label{preliminary}
\subsection{Price Models} \label{subsec-models}
%
%
Let $G=(V,E,\{w_{j}\})$ be a reduced instance of the item pricing problem. 
For a hyperedge~$e_{j} \in E$~and~a~price 
vector $\bm{p}=(p_{1},p_{2},\ldots, p_{n})$ over  
the $n$ vertices, let 
$p(e_{j})=\sum_{i \in e_{j}} p_{i}$ be the~sum~of~the~profit~on~$e_{j} \in E$,~i.e., 
the profit that is returned from the customer $e_{j} \in E$ for the price vector 
\bm{p}. 

In most of the previous works \cite{BB,BK06,BK07,Getal}, the item pricing problem 
is considered under~the~model~that $p_{i} \geq 0$ for each item $i \in V$. 
By introducing the notion of {\it loss-leader\/}, however, 
Balcan,~et al.~\cite{BBCH} considered several price models 
in which $p_{i} < 0$ for some item $i \in V$, and showed that~the~seller~could 
get more profit in the case that $p_{i} <0$ is allowed than in the case that 
$p_{i}<0$ is not allowed.~In~the~fol\-lowing, we formally present the 
definitions of price models \cite{BBCH} with respect to the reduced instance. 
\begin{df}[{\sf Positive Price Model}] \label{df-positive}
Under the condition that $p_{i}\geq 0$~for~each $i \in V$, 
find~a~price vector $\bm{p}=(p_{1},p_{2},\ldots, p_{n})$ that maximizes 
$\msc{Profit}_{\rm pos}(\bm{p}) = \sum_{e_{j} \in E: w_{j} \geq p(e_{j})} p(e_{j})$. 
%
%
%
%
\end{df}
\begin{df}[{\sf Discount Model}] \label{df-discount}
Find a price vector $\bm{p}=(p_{1},p_{2},\ldots, p_{n})$ that maximizes 
$\msc{Profit}_{\rm disc}(\bm{p}) = \sum_{e_{j} \in E: w_{j} \geq p(e_{j})} p(e_{j})$. 
%
%
%
%
\end{df}
\begin{df}[{\sf $\bm{B}$-Bounded Discount Model}] \label{df-B-discount}
Under the conditions that $d_{i}=B$ and $p_{i} \geq -B$~for~each~$i \in V$, 
find a price vector $\bm{p}=(p_{1},p_{2},\ldots, p_{n})$ that maximizes 
$\msc{Profit}_{B}(\bm{p}) = \sum_{e_{j} \in E: w_{j} \geq p(e_{j})} p(e_{j})$. 
%
%
%
%
\end{df}
\begin{df}[{\sf Coupon Model}] \label{df-coupon}
Find a price vector $\bm{p}=(p_{1},p_{2},\ldots, p_{n})$ that maximizes 
$\msc{Profit}_{\rm coup}(\bm{p}) = \sum_{e_{j} \in E: w_{j} \geq p(e_{j})} 
\max\{p(e_{j}),0\}$. 
%
%
%
%
\end{df}
Under the coupon model, if $w_{j}\leq 0$, then $e_{j}$ never contributes to 
the profit for any price vector~$\bm{p}$.~So~without loss of generality, 
we assume that $w_{j} > 0$ for each $j \in [1,m]$ under the coupon model. 
%
\subsection{Highway Problem} \label{subsec-highway-problem}
%
For any pair of integers $a \leq b$, let $[a,b]=\{a,a+1,\ldots,b\}$. 
Informally, we say that $G=(V,E,\{w_{j}\})$~is~an instance 
of the {\it line\/} highway problem \cite{Getal} 
if each $e_{j} \in E$ is an interval in the 
line~on~$V$. We introduce the {\it cycle\/} highway problem as a generalization 
of the line highway problem, and we~say~that~$G=(V,E,\{w_{j}\})$~is an 
instance of the cycle highway problem if 
each $e_{j} \in E$ is an interval~in~the~cycle on $V$. 
%
\begin{df}
\label{df-line-highway}
We say that $G=(V,E,\{w_{j}\})$ 
is a reduced instance of 
the {\sf line} highway problem~if $e_{j}=[j_{s},j_{t}] \subseteq V$ 
for each $e_{j} \in E$, 
where $V=[1,n]$ and $1 \leq j_{s} \leq j_{t} \leq n$. 
\end{df}
\begin{df}
\label{df-cycle-highway}
We say that $G=(V,E,\{w_{j}\})$ 
is a reduced instance of the {\sf cycle} highway 
problem~if~$e_{j}=[j_{s},j_{t}] \subseteq V$ or 
$e_{j}=[j_{t},n] \cup [1,j_{s}] \subseteq V$ 
for each $e_{j} \in E$, 
where $V=[1,n]$ and $1 \leq j_{s} \leq j_{t} \leq n$. 
\end{df}
We say that $G=(V,E,\{w_{j}\})$ is an instance of 
the line (or cycle) highway problem with~{\it $[s,\ell]$-valuation\/}
if $s = \min_{j \in [1,m]} w_{j}$ and $\ell = \max_{j \in [1,m]} w_{j}$. 
In particular, we say that $G=(V,E,\{w_{j}\})$ is an instance of 
the line (or cycle) highway problem with a {\it single\/} valuation 
if $w_{j}=w>0$ for each~$j \in [1,m]$. 
%
\subsection{DAG Representation of the Line Highway Problem} \label{subsec-DAG}
%
In this subsection, we present the DAG representation of the line highway problem 
due to Balcan,~et~al. \cite[\S 3]{BBCH}. For a reduced instance 
$G=(V,E,\{w_{j}\})$ of the line highway problem, 
define the DAG representation $H=(U,F,\{w_{j}\})$ of $G$ as follows: 
For $V=\{1,2,\ldots, n\}$, let $U=\{u_{0},u_{1},\ldots, u_{n}\}$ be the set of 
$n+1$~vertices,  
and for each $e_{j}=[j_{s},j_{t}]\in E$, 
let $f_{j}=(u_{j_{s}-1},u_{j_{t}}) \in F$ be the 
arc $u_{j_{s}-1} \rightarrow u_{j_{t}}$ with weight $w_{j}$. 

Let $\bm{p}=(p_{1},p_{2},\ldots, p_{n})$ be a  price vector for 
$G=(V,E,\{w_{j}\})$. Then for the DAG representation~$H=(U,F,\{w_{j}\})$ of $G$, 
define the partial sum for $u_{i} \in U$ by 
$s_{i}=\sum_{h=1}^{i} p_{h}$, where $s_{0}=0$. 
On the other hand,~let $\bm{s}=(s_{0},s_{1},\ldots, s_{n})$ 
be the partial sum vector for the DAG representation $H$ of $G$. 
Then we can define the price vector 
$\bm{p}=(p_{1},p_{2},\ldots, p_{n})$ to be $p_{i}=s_{i}-s_{i-1}$ for each 
$i \in [1,n]$. 
%
\section{Algorithms for the Line Highway Problem} \label{sec-line-highway}
%
Balcan, et al. \cite[Theorem 3]{BBCH} showed a 2.33-approximation algorithm 
for the line highway problem~with~a single valuation. In this section, we consider 
the line highway problem with $[s,\ell]$-valuation. 

Let $G=(V,E,\{w_{j}\})$ be a reduced instance of the line highway 
problem with $[s,\ell]$-valuation.
For each $x \in [s,\ell]$, we use $E_{x}$ to denote 
the set of customers with valuation 
$x \in [s,\ell]$~and~let~$m_{x} =\leng{E_{x}}$. 
Note that $m_{s}+m_{s+1}+\cdots + m_{\ell}=m=\leng{E}$. 
Under the coupon model, 
let $\bm{p}_{\rm coup}^{*}$ be the price~vector~with~the~maximum profit, and let 
$\msc{Opt}_{\rm coup}(G)=\msc{Profit}_{\rm coup}(\bm{p}_{\rm coup}^{*})$  
be the maximum total profit returned from the customers in $E$. 
For each $x \in [s,\ell]$, we use $\msc{Opt}_{\rm coup}^{x}(G)$ to denote 
the fraction of 
$\msc{Opt}_{\rm coup}(G)$ that~is~returned~from~the customers in $E_{x}$ 
for the optimal price vector $\bm{p}_{\rm coup}^{*}$. 
From the definition of $\msc{Opt}_{\rm coup}^{x}(G)$ 
for each $x \in [s,\ell]$,~we immediately have that 
$\msc{Opt}_{\rm coup}(G) = \msc{Opt}_{\rm coup}^{s}(G) + 
\msc{Opt}_{\rm coup}^{s+1}(G) + \cdots + 
\msc{Opt}_{\rm coup}^{\ell}(G)$. 

Under the coupon model, our algorithm for the line highway problem with 
$[s,\ell]$-valuation consists of two algorithms {\sc Line\_Random} and {\sc Line\_Cut}. 
%
\subsection{Algorithm: L{\small INE\_}R{\small ANDOM}}
%
In this subsection, we present the algorithm {\sc Line\_Random} 
for the line highway problem with~$[s,\ell]$-valua\-tion. 
The description of the algorithm {\sc Line\_Random} is given in 
Figure \ref{fig-line-random}. 
\begin{thm} \label{thm-line-random}
On an instance $G=(V,E,\{w_{j}\})$ of the line highway problem 
with $[s,\ell]$-valuation, 
the~algo\-rithm {\sc Line\_Random} 
outputs a price vector $\bm{\sigma}$ that satisfies
\[
\frac{\msc{Opt}_{\rm coup}(G)}{{\bf E}\left[\msc{Profit}_{\rm coup}(\bm{\sigma})\right]} 
\leq \left\{
\begin{array}{cl}
3/r 
& 0 < r \leq 1/2;\\
6 
& 1/2< r \leq 1, 
\end{array} \right. 
\]
where $r=s/\ell$ is the ratio between the smallest and the largest valuations. 
\end{thm}
\begin{figure}[!htb]
\fbox{
\begin{minipage}{17.25cm}
\begin{minipage}{17.0cm}\medskip
%
%
\makebox[1.3cm][r]{{\sf Input:}} A reduced instance $G=(V,E,\{w_{j}\})$ 
of the line highway problem with $[s,\ell]$-valuation.\\
\makebox[1.3cm][r]{{\sf Output:}} A price vector 
$\bm{\sigma}=(\sigma_{1},\sigma_{2},\ldots, \sigma_{n})$ for $G$.
\begin{namelist}{~~2.}
\item[1.] Construct the DAG representation 
$H=(U,F,\{w_{j}\})$ of $G=(V,E,\{w_{j}\})$. 
\item[2.] For each $u_{i} \in U$, assign a partial sum $s_{i} \in [0,\ell]$ 
for $u_{i}$ uniformly and at random. 
\item[3.] For each $i \in V$, compute a price $\sigma_{i}=s_{i}-s_{i-1}$ 
for the item $i$ 
and let $\bm{\sigma}=(\sigma_{1},\sigma_{2},\ldots, \sigma_{n})$. 
\end{namelist}
\end{minipage}
\vspace*{0.25cm}
\end{minipage}}
%
%
\caption{The algorithm {\sc Line\_Random}} \label{fig-line-random}
\end{figure}
%
%
{\bf Proof:} We begin by showing the following claims. 
%
\begin{claim} \label{claim-opt-random}
$\msc{Opt}_{\rm coup}^{x}(G) \leq m_{x} \cdot x$ for each $x \in [s,\ell]$. 
\end{claim}
{\bf Proof:} For each $x \in [s,\ell]$, the maximum profit returned from a 
customer $e_{j} \in E_{x}$ is at most $x$. \QED
\begin{claim} \label{claim-line-random}
For each $x \in [s,\ell]$, let ${\bf E}[\msc{Profit}_{\rm coup}^{x}(\bm{\sigma})]$ 
be the expected profit returned from~the~set~$E_{x}$~of~customers 
by the algorithm {\sc Line\_Random}. Then 
\[
{\bf E}\left[\msc{Profit}_{\rm coup}^{x}(\bm{\sigma})\right] 
= \frac{m_{x}}{6(\ell+1)^{2}}\cdot x(x+1)(-2x+3\ell+2).
\]
\end{claim}
{\bf Proof:} For each $x \in [s,\ell]$ and each $e \in E_{x}$, 
let $Y_{x}^{e}$ be the profit returned from a customer~$e \in E_{x}$~in~Step~2 
of the algorithm {\sc Line\_Random}, 
and let $Y_{x}=\sum_{e \in E_{x}} Y_{x}^{e}$. 
For each $e \in E_{x}$, we estimate~${\bf E}[Y_{x}^{e}]$. 
\begin{eqnarray*}
{\bf E}[Y_{x}^{e}] & = & \frac{1}{(\ell+1)^{2}} 
\left\{1 \cdot \ell + 2\cdot (\ell-1)+\cdots + x \cdot (\ell-x+1)\right\}\\
& = & \frac{1}{(\ell+1)^{2}} \sum_{k=1}^{x} k \cdot (\ell-k+1) 
= \frac{1}{6 (\ell+1)^{2}} \cdot x(x+1)(-2x + 3\ell+2).
\end{eqnarray*}
Thus from the linearity of expectation \cite{MR} and the fact that 
$m_{x}=\leng{E_{x}}$, it follows that for~each~$x \in [s,\ell]$, 
${\bf E}[\msc{Profit}_{\rm coup}^{x}(\bm{\sigma})] = {\bf E}[Y_{x}]=
\sum_{e \in E_{x}} {\bf E}[Y_{x}^{e}]=
\leng{E_{x}} \cdot {\bf E}[Y_{x}^{e}]
=m_{x} \cdot {\bf E}[Y_{x}^{e}]$. \QED\medskip
%
%

From Claims \ref{claim-opt-random} and \ref{claim-line-random}, 
it follows that for each $x \in [s,\ell]$, 
\[
\frac{{\bf E}[\msc{Profit}_{\rm coup}^{x}(\bm{\sigma})]}{\msc{Opt}_{\rm coup}^{x}(G)} 
\geq \frac{1}{6(\ell+1)^{2}} \cdot (x+1)(-2x + 3\ell+2).
\]
Let $f(x)=(x+1)(-2x + 3\ell+2)$ and let $F_{min} = \min_{x \in [s,\ell]} f(x)$. 
Since the function $f$ is convex~with~re\-spect to $x \in [s,\ell]$, 
we have that $F_{min}= \min\{f(s),f(\ell)\}$. Let 
$g(s,\ell)=f(\ell)-f(s) = (\ell-2s)(\ell-s)$~and~this~implies that 
$f(\ell) \geq f(s)$ if $\ell \geq 2s$; $f(\ell) < f(s)$ if $s \leq \ell < 2s$. 

For the case that $\ell \geq 2s$, it follows that $F_{min}=f(s)$. 
So we have that for each $x \in [s,\ell]$, 
\begin{eqnarray*}
\frac{{\bf E}[\msc{Profit}_{\rm coup}^{x}(\bm{\sigma})]}{\msc{Opt}_{\rm coup}^{x}(G)} 
& \geq & 
\frac{f(x)}{6(\ell+1)^{2}} \geq \frac{f(s)}{6(\ell+1)^{2}} = 
\frac{(s+1)(3\ell-2s+2)}{6(\ell+1)^{2}}\\
& = & \frac{1}{6} \cdot \frac{s+1}{\ell+1} \cdot \frac{3\ell-2s+2}{\ell+1}
\geq \frac{1}{6} \cdot \frac{s}{\ell} \cdot 
\left( 3 - \frac{2s+1}{\ell+1}\right)\\
& \geq & \frac{1}{6} \cdot \frac{s}{\ell} \cdot \left(3 - \frac{\ell+1}{\ell+1}\right)
=\frac{1}{3}\cdot \frac{s}{\ell} = \frac{r}{3},  
\end{eqnarray*} 
which implies that ${\bf E}[\msc{Profit}_{\rm coup}^{x}(\bm{\sigma})]
\geq (r/3) \cdot \msc{Opt}_{\rm coup}^{x}(G)$ for each $x \in [s,\ell]$. 
Thus we have that 
\begin{eqnarray*}
\frac{{\bf E}[\msc{Profit}_{\rm coup}(\bm{\sigma})]}{\msc{Opt}_{\rm coup}(G)} & = & 
\frac{{\bf E}[\msc{Profit}_{\rm coup}^{s}(\bm{\sigma})]+
{\bf E}[\msc{Profit}_{\rm coup}^{s+1}(\bm{\sigma})]+ \cdots + 
{\bf E}[\msc{Profit}_{\rm coup}^{\ell}(\bm{\sigma})]}{
\msc{Opt}_{\rm coup}^{s}(G)+\msc{Opt}_{\rm coup}^{s+1}(G)+\cdots + 
\msc{Opt}_{\rm coup}^{\ell}(G)}\\
& \geq & \frac{r}{3} \cdot 
\frac{\msc{Opt}_{\rm coup}^{s}(G)+\msc{Opt}_{\rm coup}^{s+1}(G)+\cdots + 
\msc{Opt}_{\rm coup}^{\ell}(G)}{
\msc{Opt}_{\rm coup}^{s}(G)+\msc{Opt}_{\rm coup}^{s+1}(G)+\cdots + 
\msc{Opt}_{\rm coup}^{\ell}(G)} = \frac{r}{3}. 
\end{eqnarray*}
For the case that $s \leq \ell < 2s$, it follows that $F_{min}=f(\ell)$. 
So we have that for each $x \in [s,\ell]$, 
\[
\frac{{\bf E}[\msc{Profit}_{\rm coup}^{x}(\bm{\sigma})]}{\msc{Opt}_{\rm coup}^{x}(G)} 
\geq 
\frac{f(x)}{6(\ell+1)^{2}} \geq \frac{f(\ell)}{6(\ell+1)^{2}} = 
\frac{(\ell+1)(\ell+2)}{6(\ell+1)^{2}}
= \frac{1}{6} \cdot \frac{\ell+2}{\ell+1} > \frac{1}{6}, 
\]
which implies that ${\bf E}[\msc{Profit}_{\rm coup}^{x}(\bm{\sigma})]
\geq (1/6) \cdot \msc{Opt}_{\rm coup}^{x}(G)$ for each $x \in [s,\ell]$. 
Thus~in~a~way~similar~to~the above, 
we have that ${\bf E}[\msc{Profit}_{\rm coup}(\bm{\sigma}) 
\geq (1/6) \cdot \msc{Opt}_{\rm coup}(G)$. \BQED
%
\subsection{Algorithm: L{\small INE\_}C{\small UT}} \label{subsec-line-cut}
\begin{figure}[htb]
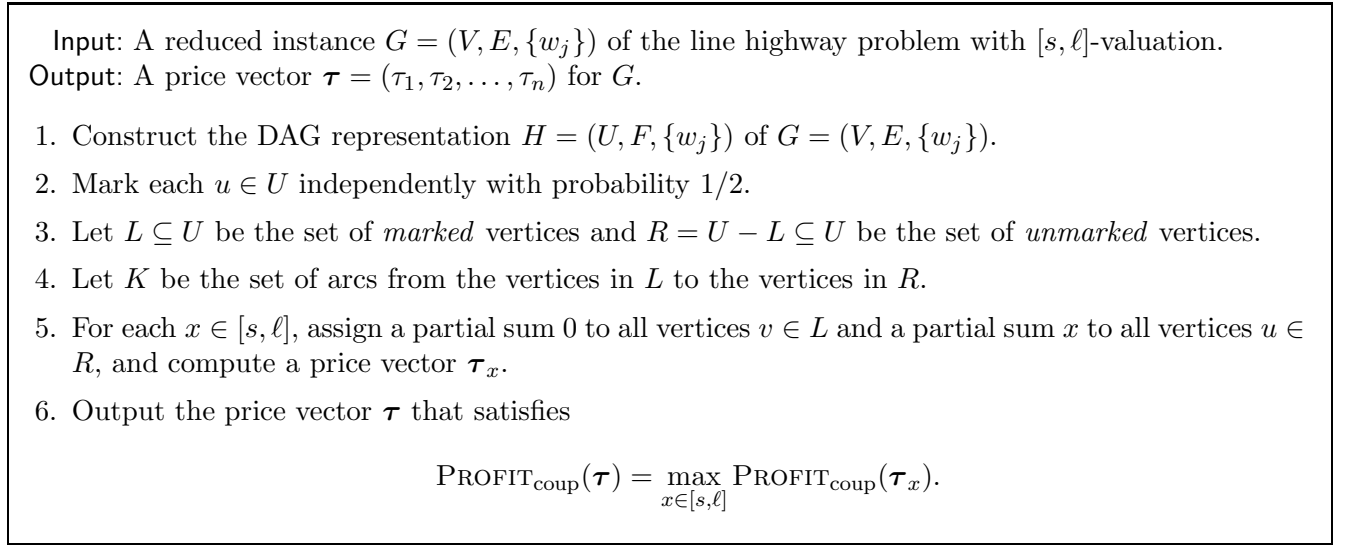

%
%
\fbox{
\begin{minipage}{17.25cm}
\begin{minipage}{17.0cm}\medskip
%
%
\makebox[1.3cm][r]{{\sf Input:}} A reduced instance $G=(V,E,\{w_{j}\})$ 
of the line highway problem with $[s,\ell]$-valuation.\\
\makebox[1.3cm][r]{{\sf Output:}} A price vector 
$\bm{\tau}=(\tau_{1},\tau_{2},\ldots, \tau_{n})$ for $G$.
\begin{namelist}{~~2.}
\item[1.] Construct the DAG representation 
$H=(U,F,\{w_{j}\})$ of $G=(V,E,\{w_{j}\})$. 
\item[2.] Mark each $u \in U$ independently with probability $1/2$. 
\item[3.] Let $L \subseteq U$ be the set of {\it marked\/} vertices 
and $R=U-L \subseteq U$ be the set of {\it unmarked\/} vertices. 
\item[4.] Let $K$ be the set of arcs from the vertices in $L$ to 
the vertices in $R$. 
%
\item[5.] For each~$x \in [s,\ell]$, assign a partial sum $0$ to 
all vertices $v \in L$ and a partial sum $x$~to~all~vertices~$u \in R$, 
and compute a price vector $\bm{\tau}_{x}$. 
\item[6.] Output the price vector $\bm{\tau}$ that satisfies 
\[
\msc{Profit}_{\rm coup}(\bm{\tau}) = \max_{x \in [s,\ell]} 
\msc{Profit}_{\rm coup}(\bm{\tau}_{x}). 
\]
%
%
%
%
%
%
%
%
%
%
%
%
%
%
\end{namelist}
\end{minipage}\\[0.25cm]
\end{minipage}}
%
%
%
\caption{The Algorithm {\sc Line\_Cut}} \label{fig-line-cut}
\end{figure}

%
In this subsection, we present the algorithm {\sc Line\_Cut} 
for the line highway problem~with~$[s,\ell]$-val\-uation. 
The description of the algorithm {\sc Line\_Cut} is given in Figure 
\ref{fig-line-cut}. 
\begin{thm} \label{thm-simple-line-cut}
On an instance $G=(V,E,\{w_{j}\})$ of the line highway problem 
with $[s,\ell]$-valuation,~the~algo\-rithm {\sc Line\_Cut} 
outputs a price vector $\bm{\tau}$ that satisfies
\[
\frac{\msc{Opt}_{\rm coup}(G)}{{\bf E}[\msc{Profit}_{\rm coup}(\bm{\tau})]} \leq 
4(1-\ln r), 
\]
where $r=s/\ell$ is the ratio between the smallest and the largest valuations. 
\end{thm}
{\bf Proof:} 
For the set $K$ of the arcs from the vertices in $L$ to the 
vertices in $R$, let $\msc{Val}(K)$~be~the~sum~of~the valuations of 
the arcs in $K$. For each $x \in [s,\ell]$, 
let $K_{x}$ be the set of~arcs~in~$K$~with~valua\-tion 
$x$~and~let~$m_{x}= \leng{K_{x}}$. Then it is immediate to see that 
$\msc{Val}(K) = m_{s}\cdot s+ m_{s+1}\cdot (s+1) + \cdots + m_{\ell}\cdot \ell$. To  
%
%
%
%
comlete~the~proof of the theorem, we need to show the following claims: 
%
\begin{claim} \label{claim-cut-val}
${\bf E}[\msc{Val}(K)] = (1/4) \cdot \sum_{f_{j} \in F} w_{j} 
= (1/4)\cdot \sum_{e_{j} \in E} w_{j} 
\geq (1/4) \cdot \msc{Opt}_{\rm coup}(G)$. 
\end{claim}
{\bf Proof:} The first equality follows from the definition of $U$ and the first 
inequality is trivial. \QED
\begin{claim} \label{claim-simple-line-cut-profit}
For each $x \in [s,\ell]$, the following holds$:$
%
\[
\msc{Val}(K)=\msc{Profit}_{\rm coup}(\bm{\tau}_{s})
+ \sum_{x=s+1}^{\ell} \frac{\msc{Profit}_{\rm coup}(\bm{\tau}_{x})}{x}. 
\]
\end{claim}
{\bf Proof:} From the definition of $\bm{\tau}_{x}$, 
we have that $\msc{Profit}_{\rm coup}(\bm{\tau}_{x}) 
= m_{x}\cdot x+m_{x+1}\cdot x+ \cdots + m_{\ell}\cdot x$~for~each $x \in [s,\ell]$. 
%
%
%
%
Then the claim immediately follows from the definition of $\msc{Val}(K)$. \QED\medskip
%
%
%
%
%

From Claims \ref{claim-cut-val} and \ref{claim-simple-line-cut-profit} and the 
definition of $\bm{\tau}$, it follows that 
\begin{eqnarray*}
\lefteqn{\frac{1}{4}\cdot \msc{Opt}_{\rm coup}(G) \leq 
{\bf E}[\msc{Val}(K)]
= 
{\bf E}[\msc{Profit}_{\rm coup}(\bm{\tau}_{s})]  
+ \sum_{x=s+1}^{\ell} \frac{{\bf E}[\msc{Profit}_{\rm coup}(\bm{\tau}_{x})]}{x}}\\
& \leq & {\bf E}[\msc{Profit}_{\rm coup}(\bm{\tau})] + 
\sum_{x=s+1}^{\ell} 
\frac{{\bf E}[\msc{Profit}_{\rm coup}(\bm{\tau})]}{x}
= \left(1 + \sum_{k=s+1}^{\ell} \frac{1}{k}\right) \cdot 
{\bf E}[\msc{Profit}_{\rm coup}(\bm{\tau})]. 
\end{eqnarray*}
Since $\sum_{k=s+1}^{\ell} 1/k \leq \ln (\ell/s) = -\ln r$, we have that 
$\msc{Opt}_{\rm coup}(G) \leq 4(1-\ln r) \cdot 
{\bf E}[\msc{Profit}_{\rm coup}(\bm{\tau})]$. \BQED
\begin{rem} \label{remark-derandomize}
The algorithm {\sc Line\_Cut} can be easily derandomized by applying 
pairwise~independent $0/1$-random variables with a small sample space 
{\rm \cite{LW}} in {\rm Step 2}. 
\end{rem}
%
\subsection{Algorithm: $\bm{\mbox{L{\normalsize INE}}_{[s,\ell]}}$}
\label{subsec-line-[s,l]}
%
The algorithm $\msc{Line}_{[s,\ell]}$ works as follows: 
On an instance $G=(V,E,\{w_{j}\})$ of the line highway~problem~with 
$[s,\ell]$-valuation, 
(1) run {\sc Line\_Random} on $G$ to get the price vector $\bm{\sigma}$; 
(2) run {\sc Line\_Cut} on $G$ to get the price vector $\bm{\tau}$; 
(3) output the price vector $\bm{p}$ that satisfies 
\[
\msc{Profit}_{\rm coup}(\bm{p}) = \max\left\{
\msc{Profit}_{\rm coup}(\bm{\sigma}), \msc{Profit}_{\rm coup}(\bm{\tau})\right\}. 
\]
From Theorems \ref{thm-line-random} and \ref{thm-simple-line-cut}, 
we immediately have the following theorem: 
\begin{thm} \label{thm-line-highway}
On an instance $G=(V,E,\{w_{j}\})$ of the line highway problem 
with $[s,\ell]$-valuation,~the~algo\-rithm $\msc{Line}_{[s,\ell]}$ 
outputs a price vector $\bm{p}$ that satisfies
\[
\frac{\msc{Opt}_{\rm coup}(G)}{{\bf E}[\msc{Profit}_{\rm coup}(\bm{p})]} 
\leq \left\{
\begin{array}{cl}
4(1-\ln r) & 0 \leq r \leq \alpha,~1/\sqrt{e} \leq r \leq 1;\\
3/r & \alpha < r \leq 1/2;\\
6 & 1/2 < r < 1/\sqrt{e}, 
\end{array} \right.
\]
where $r=s/\ell$ is the ratio between the smallest and the largest valuations 
and $\alpha \approx 0.3824$,~i.e.,~$\alpha$~is~the~solution 
of the equality $3/x =4(1-\ln x)$. 
\end{thm}
%
\section{Algorithms for the Cycle Highway Problem} \label{cycle-highway}
%
In this section, we first consider the cycle highway problems with 
$[s,\ell]$-valuation and~then~we~consider~the 
cycle highway problems with a single valuation as the special case of 
the cycle highway problem~with~$[s,\ell]$-valuation such that $s=\ell$, which also 
can be regarded as an extension of the line highway~problem~with~a single valuation 
discussed by Balcan, et al. \cite{BBCH}. 
%
\subsection{Algorithms for the Cycle Highway Problem with \bm{[s,\ell]}-Valuation} 
\label{subsec-cycle-highway-[s,l]}
%
In this subsection, we present an algorithm $\msc{\sc Cycle}_{[s,\ell]}$ for the 
cycle highway problem~with~$[s,\ell]$-val\-uation. 

For a reduced instance $G=(V,E,\{w_{j}\})$ of the cycle highway problem with 
$[s,\ell]$-valuation,~define~a~directed graph $H=(U,F,\{w_{j}\})$ 
as follows:  Let $V=\{1,2,\ldots, n\}$ and each item $i \in V$ 
is arranged~in~a~clockwise manner, i.e., we arrange 
$1 \rightarrow 2 \rightarrow \cdots \rightarrow n \rightarrow 1$. 
For~each~$e_{j} \in E$,~let~$1 \leq j_{s} \leq j_{t} \leq n$.~If~$e_{j}=[j_{s},j_{t}] \subseteq V$, then 
we define $f_{j}=(j_{s},j_{t})$ to be an arc $j_{s} \rightarrow j_{t}$ 
with valuation $w_{j}$; 
if~$e_{j}=[j_{t},n] \cup [1,j_{s}]$,~then~we~define 
$f_{j}=(j_{t},j_{s})$ to be an arc $j_{t} \rightarrow j_{s}$ 
with valuation~$w_{j}$.~Let~$F=\{f_{j}: e_{j} \in E\}$~be~the~set~of~arcs~and~let  
$U=\{j_{s} \in V:e_{j} =(j_{s},j_{t})\in E\} \cup 
\{j_{t} \in V: e_{j}=(j_{s},j_{t}) \in E\}$ be the set of vertices. 

The description of the algorithm $\msc{\sc Cycle}_{[s,\ell]}$ is given in 
Figure \ref{fig-cycle-[s,l]}.
%
%
\begin{figure}[!h]
%
%
\fbox{
\begin{minipage}{17.25cm}
\begin{minipage}{17.0cm}\medskip
%
%
\makebox[1.3cm][r]{{\sf Input:}} A reduced instance $G=(V,E,\{w_{j}\})$ 
of the cycle highway problem with $[s,\ell]$-valuation.\\
\makebox[1.3cm][r]{{\sf Output:}} A price vector 
$\bm{p}=(p_{1},p_{2},\ldots, p_{n})$ for $G$.
\begin{namelist}{~~2.}
\item[1.] For $G=(V,E,\{w_{j}\})$, construct a directed graph $H=(U,F,\{w_{j}\})$. 
\item[2.] Mark each $u \in U$ independently with probability 1/2. 
\item[3.] Let $L \subseteq U$ be the set of {\it  marked\/} vertices 
and $R=U-L \subseteq U$  be the set of {\it unmarked\/} vertices. 
\item[4.] Let $J_{H}=\{f_{j}= (a_{j},b_{j}) \in F: a_{j} \in L, b_{j} \in R\}$ 
be the set of arcs from the vertices~in~$L$~to~the~vertices in $R$ 
and remove all arcs in $F-J_{H}$. 
\item[5.] For each $x \in [s,\ell]$, assign a partial sum $-x/2$ to all vertices 
$v \in L$ and a partial~sum~$x/2$~to~all~vertices $u \in R$, and compute a 
price vector $\bm{p}_{x}$. 
\item[6.] Output the price vector $\bm{p}$ that satisfies 
$\msc{Profit}_{\rm coup}(\bm{p}) 
= \max_{x \in [s,\ell]} \msc{Profit}_{\rm coup}(\bm{p}_{x})$. 
\end{namelist}
\end{minipage}\\[0.25cm]
\end{minipage}}
%
%
%
\caption{The Algorithm $\msc{Cycle}_{[s,\ell]}$} \label{fig-cycle-[s,l]}
\end{figure}
\begin{thm} \label{thm-cycle-[s,l]}
On an instance $G=(V,E,\{w_{j}\})$ of the cycle highway problem 
with $[s,\ell]$-valuation,~the~al\-gorithm $\msc{Cycle}_{[s,\ell]}$ 
outputs a price vector $\bm{p}$ that satisfies
\[
\msc{Opt}_{\rm coup}(G) \leq 4(1-\ln r) \cdot {\bf E}[\msc{Profit}_{\rm coup}(\bm{p})], 
\]
where $r=s/\ell$ is the ratio between the smallest and the largest valuations. 
\end{thm}
{\bf Proof:} For the set $J_{H}$ of arcs 
from the vertices in $L$ to the vertices in $R$, 
let $\msc{Val}(J_{H})$~be~the~sum~of~the~valuations of the 
arcs in $J_{H}$. For each $x \in [s,\ell]$, let $J_{H}^{x}$ be the set of 
arcs~in~$J_{H}$~with valuation~$x$~and~let~$m_{x}=\leng{J_{H}^{x}}$. 
Then we~can~show~the~following claims: 
\begin{claim} \label{claim-cycle-K}
${\bf E}[\msc{Val}(J_{H})] = (1/4)\cdot \sum_{f_{j} \in F} w_{j}= 
(1/4) \cdot \sum_{e_{j} \in E} w_{j} \geq (1/4) \cdot \msc{Opt}_{\rm coup}(G)$. 
\end{claim}
{\bf Proof:} This can be shown in a way similar to the proof of 
Claim \ref{claim-cut-val}. \QED
\begin{claim} \label{claim-cycle-x}
For the price vector $\bm{p}_{s},\bm{p}_{s+1},\ldots, 
\bm{p}_{\ell}$ and $\msc{Val}(J_{H})$, the following holds$:$
\[
\msc{Val}(J_{H}) = \msc{Profit}_{\rm coup}(\bm{p}_{s}) + \sum_{x=s+1}^{\ell} 
\frac{\msc{Profit}_{\rm coup}(\bm{p}_{x})}{x}.
\]
\end{claim}
{\bf Proof:} Note that each vertex $v \in L$ has no incoming arcs 
and each vertex $u \in R$ has~no~outgoing~arcs.~Define 
the set $J_{G}$ of intervals to be 
$e_{j} \in J_{G}$ if $f_{j} \in J_{H}$. 
Thus for each $x \in [s,\ell]$, assigning a partial~sum~$-x/2$~to all 
vertices $v \in L$ and assigning a partial sum $x/2$ to all vertices 
$u \in R$ implies that each $e_{j} \in J_{G}$~is assigned $x=x/2-(-x/2)$ 
as a total sum of prices for the corresponding items in $e_{j}$. 
To~define~a~price~vector~$\bm{p}_{x}$,~we appropriately assign prices 
to all items that are not assigned prices (this does not reduce the profit returned 
from the customers $e_{j} \in J_{G}$). 
Then we have that 
$\msc{Profit}_{\rm coup}(\bm{p}_{x})=m_{x}\cdot x+m_{x+1}\cdot x+\cdots + 
m_{\ell}\cdot x$~for~each $x \in [s,\ell]$.
In a way similar to the 
proof of Claim \ref{claim-simple-line-cut-profit}, the claim follows 
from the definition of $\msc{Val}(J_{H})$,~i.e., 
$\msc{Val}(J_{H})=m_{s}\cdot s+m_{s+1}\cdot (s+1)+\cdots +m_{\ell} \cdot \ell$. 
\QED\medskip

From Claims \ref{claim-cycle-K} and \ref{claim-cycle-x} and the definition of 
$\bm{p}$, it follows that 
\begin{eqnarray*}
\lefteqn{\frac{1}{4}\cdot \msc{Opt}_{\rm coup}(G) \leq 
{\bf E}[\msc{Val}(J_{H})]
= 
{\bf E}[\msc{Profit}_{\rm coup}(\bm{p}_{s})]  
+ \sum_{x=s+1}^{\ell} \frac{{\bf E}[\msc{Profit}_{\rm coup}(\bm{p}_{x})]}{x}}\\
& \leq & {\bf E}[\msc{Profit}_{\rm coup}(\bm{p})] + 
\sum_{x=s+1}^{\ell} 
\frac{{\bf E}[\msc{Profit}_{\rm coup}(\bm{p})]}{x}
= \left(1 + \sum_{k=s+1}^{\ell} \frac{1}{k}\right) \cdot 
{\bf E}[\msc{Profit}_{\rm coup}(\bm{p})]. 
\end{eqnarray*}
Since $\sum_{k=s+1}^{\ell} 1/k \leq \ln (\ell/s) = -\ln r$, we have that 
$\msc{Opt}_{\rm coup}(G) \leq 4(1-\ln r) \cdot 
{\bf E}[\msc{Profit}_{\rm coup}(\bm{p})]$. \BQED
\begin{rem} \label{remark-derandomize-1}
The algorithm $\msc{Cycle}_{[s,\ell]}$ can be easily derandomized by applying 
pairwise~independent~$0/1$-random variables with a small sample space 
{\rm \cite{LW}} in {\rm Step 2}. 
\end{rem}
%
%
\subsection{Algorithms for the Cycle Highway Problem with a Single Valuation} 
\label{subsec-cycle-highway-single}
%
Let $G=(V,E,\{w_{j}\})$ be a reduced instance of the cycle highway problem with a 
single valuation,~i.e.,~$r=1$. So it follows 
from Theorem \ref{thm-cycle-[s,l]} that 
$\msc{Opt}_{\rm coup}(G) \leq 4 \cdot \msc{Profit}_{\rm coup}(\bm{p})$. 
To improve this,~we~present~the~algorithm 
{\sc Cyc\_Single\_Val} for the cycle highway problem with a single valuation. 
Without loss of generality, we assume that $w_{j}=1$ for 
each customer $e_{j}\in E$ and we~use~$G=(V,E)$ to denote~an~instance~of~the~cycle 
highway problem with a single valuation. 
The algorithm {\sc Cyc\_Single\_Val} is given in 
Figure \ref{fig-cyc-single-val}. 
%
%
\begin{figure}[htb]
%
%
\fbox{
\begin{minipage}{17.25cm}
\begin{minipage}{17.0cm}\medskip
%
%
\makebox[1.3cm][r]{{\sf Input:}} A reduced instance $G=(V,E)$ 
of the cycle highway problem with a single valuation.\\
\makebox[1.3cm][r]{{\sf Output:}} A price vector 
$\bm{p}=(p_{1},p_{2},\ldots, p_{n})$ for $G$.
\begin{namelist}{~~2.}
\item[1.] Choose an item $h \in V$ arbitrarily. 
\item[2.] Let $J_{\rm in}= \{j \in [1,m]: e_{j} \in E, h \in e_{j} \}$ and  
$E_{\rm in} = \{e_{j} \in E: i \in J_{\rm in}\}$. Let 
$V_{\rm in}=\cup_{j \in J_{\rm in}} e_{j}$~be~the set of items that 
the customers in $E_{\rm in}$ are interested in. 
\item[3.] Let $J_{\rm out}= \{j \in [1,m]: e_{j} \in E, h \not \in e_{j} \}$ and   
$E_{\rm out}=\{e_{j} \in E: j \in J_{\rm out}\}$. Let 
$V_{\rm out}=\cup_{j \in J_{\rm out}} e_{j}$~be~the set of items that 
the customers in $E_{\rm out}$ are interested in. 
\item[4.] Define a price vector $\bm{\sigma}$ by 
assigning $x$ to $1$ and by assigning $0$~to~all~$i \in V-\{h\}$.
\item[5.] Regard $G_{\rm out}=(V_{\rm out},E_{\rm out})$ as an instance of 
the line highway problem with a single valuation. 
\item[6.] On input $G_{\rm out}=(V_{\rm out},E_{\rm out})$, run the algorithm due to 
Balcan,~et al. \cite[Theorem 3]{BBCH}~to~compute~a price vector $\bm{\tau}_{\rm out}$ 
for the set $V_{\rm out}$ of items. 
\item[7.] For each $x \in \{-1,0,1,2\}$, define a price vector 
$\bm{\tau}_{\rm in}^{x}$ for the set $V_{\rm in}$ of items by assigning~$x$~to~$h$ and 
by assigning $0$ to all $i \in V_{\rm in}-(V_{\rm out}\cup\{h\})$, and 
let $\bm{\tau} = (\bm{\tau}_{\rm out}, \bm{\tau}_{\rm in})$, where 
\[
\msc{Profit}_{\rm coup}(\bm{\tau}_{\rm in}) = \max_{x \in \{-1,0,1,2\}} 
\msc{Profit}_{\rm coup}(\bm{\tau}_{\rm in}^{x}). 
\]
%
%
%
%
%
\item[8.] Output the price vector $\bm{p}$ that satisfies 
\[
\msc{Profit}_{\rm coup}(\bm{p}) = \max\left\{
\msc{Profit}_{\rm coup}(\bm{\sigma}), \msc{Profit}_{\rm coup}(\bm{\tau})\right\}.
\]
\end{namelist}
\end{minipage}\\[0.25cm]
\end{minipage}}
%
%
%
\caption{The Algorithm {\sc Cyc\_Single\_Val}} \label{fig-cyc-single-val}
\end{figure}
\begin{thm} \label{thm-cycle-single-val}
On an instance $G=(V,E)$ of the cycle highway problem 
with a single valuation,~the~algo\-rithm {\sc Cyc\_Single\_Val} 
outputs a price vector $\bm{p}$ that satisfies
\[
\msc{Opt}_{\rm coup}(G) \leq 2.747 \cdot \msc{Profit}_{\rm coup}(\bm{p}).
\]
\end{thm}
{\bf Proof:} As in Subsection \ref{subsec-cycle-highway-single}, 
it is obvious that $E_{\rm in} \cup E_{\rm out} \subseteq V$. 
Without loss of generality,~we~assume~that 
$E_{\rm in} \cup E_{\rm out} = V$ (otherwise the 
instance $G$ of the cycle highway problem with a single 
valuation~can~be~regarded as an instance of the line highway problem 
with a single valuation, which 
has a 2.33-approximation algorithm due to Balcan, et al \cite[Theorem 3]{BBCH}). 
Let $\bm{p}_{\rm coup}^{*}$ be the price vector with the maximum 
profit~and $\msc{Opt}_{\rm coup}(G)=\msc{Profit}_{\rm coup}(\bm{p}_{\rm coup}^{*})$ 
be the maximum profit returned from the customers in $E$. 
For~the~optimal price vector $\bm{p}_{\rm coup}^{*}$, we use 
$\msc{Opt}_{\rm coup}^{\rm in}(G)$ 
to denote the fraction of $\msc{Opt}_{\rm coup}(G)$ that~are~returned~from~the 
customers in $E_{\rm in}$, and we also use 
$\msc{Opt}_{\rm coup}^{\rm out}(G)$ to denote 
the fraction of $\msc{Opt}_{\rm coup}(G)$ that~are~returned~from the 
customers in $E_{\rm out}$.  
It is obvious~that~$\msc{Opt}_{\rm coup}(G)=\msc{Opt}_{\rm coup}^{\rm in}(G)+
\msc{Opt}_{\rm coup}^{\rm out}(G)$. 

For each~$e_{j} \in E_{\rm in}$, let 
$e_{j}^{L}\subseteq e_{j} -\{h\}$ 
(resp. $e_{j}^{R} \subseteq e_{j}-\{h\}$) be the subinterval on the {\it left\/} 
(resp.~the~{\it right\/}) of $h$. For each $e_{j} \in E_{\rm in}$, we have that 
$e_{j}^{L} \cap e_{j}^{R} = \emptyset$ and 
$e_{j}=e_{j}^{L} \cup e_{j}^{R}\cup \{h\}$. 
To complete the~proof~of~the~theorem, we need to show the following claims: 
%
\begin{claim} \label{claim-cycle-single-val-1}
$\msc{Profit}_{\rm coup}(\bm{\sigma}) = \leng{E_{\rm in}} \geq 
\msc{Opt}_{\rm coup}^{\rm in}(G)$. 
\end{claim}
{\bf Proof:} This follows from the fact that each customer in $E_{\rm in}$ 
provides profit ``1.'' \QED
\begin{claim} \label{claim-cycle-single-val-2}
$\msc{Profit}_{\rm coup}(\bm{\tau}_{\rm out}) \geq \msc{Opt}_{\rm coup}^{\rm out}(G)/a$, 
where $a \approx 2.33$. 
\end{claim}
{\bf Proof:} This follows from the result due to Balcan, et al. 
\cite[Theorem 3]{BBCH}. \QED
\begin{claim} \label{claim-cycle-single-val-s}
For each $e_{j} \in E_{\rm in}$, the sum of the prices for the items in $e_{j}^{R}$ is 
either $0$~or~$1$,~and~the~sum~of~the prices for the items in $e_{j}^{L}$ is 
either $-1$, $0$, or $1$. 
\end{claim}
{\bf Proof:} For $G_{\rm out}=(V_{\rm out},E_{\rm out})$ in Step 6 of the 
algorithm {\sc Cyc\_Single\_Val}, 
let $H_{\rm out}=(U_{\rm out},F_{\rm out})$~be~the 
DAG representation of $G_{\rm out}$ and 
let $U_{\rm out}=\{u_{0},u_{1},\ldots,u_{k}\}$. 
Let $\bm{s}=(s_{0},s_{1},\ldots, s_{k})$~be~the~partial~sum~vector output by the 
algorithm due to Balcan, et al. \cite[Theorem 3]{BBCH} on $H_{\rm out}$. 
On input $H_{\rm out}$, the algorithm~\cite[Theorem 3]{BBCH} 
computes the directed cut $(U_{\rm out}^{L}:U_{\rm out}^{R})$~by~running~the~algorithm 
due to Feige and Goemans~\cite{FG} 
and defines the partial sum vector $\bm{s}$ by assigning~0~to~all 
$i \in U_{\rm out}^{L}$ and by assigning 1 to al $i \in U_{\rm out}^{R}$. 
For the DAG representation 
$H_{\rm out}=(U_{\rm out},F_{\rm out})$~of~$G_{\rm out}$,~it~is~easy~to~see
that $u_{0} \in U_{\rm out}$ has~no~incoming~arcs.~Thus if 
$u_{0} \in U_{\rm out}^{R}$, then by moving $u_{0}$~from~$U_{\rm out}^{R}$ to 
$U_{\rm out}^{L}$,~we~have~the directed cut 
$(U_{\rm out}^{L}\cup \{u_{0}\}:U_{\rm out}^{R}-\{u_{0}\}\})$~including more 
crossing arcs than the directed cut 
$(U_{\rm out}^{L}:U_{\rm out}^{R}\})$. So without loss of generality, we assume 
that $u_{0} \in U_{\rm out}^{L}$ for the directed cut 
$(U_{\rm out}^{L},U_{\rm out}^{R})$. This implies that $s_{0}=0$ and 
$s_{i} \in \{0,1\}$ for each $1 \leq i \leq k$. 

For each $e_{j} \in E_{\rm in} \subseteq V_{\rm in}$, let 
$e_{j}'=e_{j} \cap V_{\rm out}
=\{v_{1}^{\rm out},v_{2}^{\rm out},\ldots, v_{t}^{\rm out}\}$. 
So the sum of~the~prices~for~the~items 
$v_{1}^{\rm out},v_{2}^{\rm out},\ldots, v_{t}^{\rm out}$~is 
$s_{t}-s_{0}=s_{t} \in \{0,1\}$. On the other hand, 
we have assigned 0~to~all~$i \in V_{\rm in} - (V_{\rm out}\cup \{h\})$ 
in Step 7 of the algorithm {\sc Cyc\_Single\_Val}, 
which implies that for each $e_{j} \in E_{\rm in}$, 
the sum of the prices for the items $i \in e_{j} - (V_{\rm out} \cup \{h\})$ is 0. 
Thus we have that~for~each~$e_{j} \in E_{\rm in}$,~the sum of the prices 
for the items in $e_{j}^{R}$ is $s_{t} \in \{0,1\}$. 
In a way similar to $e_{j}^{R}$, we can immediately show that 
for each $e_{j} \in E_{\rm in}$,~the~sum of the prices 
for the items in $e_{j}^{L}$ is either $-1$, $0$, or $1$. \QED
\begin{claim} \label{claim-cycle-single-val-3}
$\msc{Profit}_{\rm coup}(\bm{\tau}) \geq \msc{Opt}_{\rm coup}^{\rm out}(G)/a+ 
\msc{Opt}_{\rm coup}^{\rm in}(G)/4$, where $a \approx 2.33$. 
\end{claim}
{\bf Proof:} Define ${\cal M}_{h}, {\cal M}_{L}, {\cal M}_{R}, 
{\cal M}_{LR} \subseteq E_{\rm in}$ as follows: 
\begin{eqnarray*}
{\cal M}_{h} & = & \{e_{j} \in E_{\rm in}: e_{j}^{L}=e_{j}^{R}=\emptyset\};\\
{\cal M}_{L} & = & \{e_{j} \in E_{\rm in}: e_{j}^{L} \neq \emptyset, 
e_{j}^{R}=\emptyset\};\\
{\cal M}_{R} & = & \{e_{j} \in E_{\rm in}: e_{j}^{L}=\emptyset, 
e_{j}^{R} \neq \emptyset\};\\
{\cal M}_{LR} & = & \{e_{j} \in E_{\rm in}: e_{j}^{L} \neq \emptyset, 
e_{j}^{R} \neq \emptyset\}.
\end{eqnarray*}
From~Claim~\ref{claim-cycle-single-val-s}, we 
have that for each $e_{j} \in E_{\rm in}$, the sum of the prices for the items in 
$e_{j}^{R}$~is~either~$0$~or~$1$,~and the sum of the prices for 
the items in $e_{j}^{L}$ is either $-1$, $0$, or $1$. 
For~each~$\beta \in \{-1,0,1\}$~and~each~$\gamma \in \{0,1\}$,~we further parti\-tion 
${\cal M}_{L}$, ${\cal M}_{R}$, 
and ${\cal M}_{LR}$ according to the sum of the prices. 
\begin{eqnarray*}
{\cal M}_{L}^{(\beta)} & = & \{e_{j} \in {\cal M}_{L}:
\mbox{the sum of the prices of items in $e_{j}^{L}$ is $\beta$}\};\\
{\cal M}_{R}^{(\gamma)} & = & \{e_{j} \in {\cal M}_{R}:
\mbox{the sum of the prices of items in $e_{j}^{R}$ is $\gamma$}\};\\
{\cal M}_{LR}^{(\beta,\gamma)} & = & \{e_{j} \in {\cal M}_{LR}:
\mbox{the sum of the prices of items in $e_{j}^{L}$ is $\beta$}\\
 & & ~~~~~~~~~~~~~~~~~~~~
\mbox{and the sum of the prices of items in $e_{j}^{R}$ is $\gamma$}\}. 
\end{eqnarray*}
If the price of $h \in V$ is $x \in \{-1,0,1,2\}$, then 
from the customers in $E_{\rm in}$, we can get 
\[
\msc{Profit}_{\rm coup}(\bm{\tau}_{\rm in}^{x}) = \left\{
\begin{array}{lcl}
\leng{{\cal M}_{LR}^{(1,1)}} & & x=-1;\\
\leng{{\cal M}_{L}^{(1)}} + \leng{{\cal M}_{R}^{(1)}} + 
\leng{{\cal M}_{LR}^{(0,1)}} + \leng{{\cal M}_{LR}^{(1,0)}} & & x=0;\\
\leng{{\cal M}_{h}} + \leng{{\cal M}_{L}^{(0)}} + \leng{{\cal M}_{R}^{(0)}} + 
\leng{{\cal M}_{LR}^{(-1,1)}} + \leng{{\cal M}_{LR}^{(0,0)}} & & x=1;\\
\leng{{\cal M}_{L}^{(-1)}} + \leng{{\cal M}_{LR}^{(-1,0)}} & & x=2;\\
%
\end{array}\right. 
\]
Thus in Step 7, 
we have that $\msc{Profit}_{\rm coup}(\bm{\tau}_{\rm in})\geq \leng{E_{\rm in}}/4 
\geq \msc{Opt}_{\rm coup}^{\rm in}(G)/4$, and 
it follows~from~Claim~\ref{claim-cycle-single-val-2} that 
$\msc{Profit}_{\rm coup}(\bm{\tau}) \geq \msc{Opt}_{\rm coup}^{\rm out}(G)/a+ 
\msc{Opt}_{\rm coup}^{\rm in}(G)/4$, where $a \approx 2.33$. \QED\medskip

From Claims \ref{claim-cycle-single-val-1} and \ref{claim-cycle-single-val-3}, 
we have that 
\begin{eqnarray*}
\msc{Profit}_{\rm coup}(\bm{p}) & = & 
\max\left\{ \msc{Profit}_{\rm coup}(\bm{\sigma}), 
\msc{Profit}_{\rm coup}(\bm{\tau})\right\}\\
& \geq & \max\left\{ \msc{Opt}_{\rm coup}^{\rm in}(G), 
\frac{\msc{Opt}_{\rm coup}^{\rm out}(G)}{a}+
\frac{\msc{Opt}_{\rm coup}^{\rm in}(G)}{4}\right\}\\
& \geq & \frac{4-a}{3a + 4} \cdot \msc{Opt}_{\rm coup}^{\rm in}(G) + 
\frac{4a}{3a+4} \cdot \left\{
\frac{\msc{Opt}_{\rm coup}^{\rm out}(G)}{a}+
\frac{\msc{Opt}_{\rm coup}^{\rm in}(G)}{4}\right\}\\
& = & \frac{4}{3a+4} \cdot \left\{
\msc{Opt}_{\rm coup}^{\rm in}(G) + 
\msc{Opt}_{\rm coup}^{\rm out}(G)\right\}
= \frac{4}{3a+4} \cdot \msc{Opt}_{\rm coup}(G). 
\end{eqnarray*}
Since $a \approx 2.33$, we have that the algorithm {\sc Cyc\_Single\_Val} 
is a 2.747-approximation algorithm~for the cycle highway problem with a single 
valuation. \BQED
%
\section{Concluding Remarks} \label{remarks}
%
In this paper, we have considered the line and cycle highway problems with 
$[s,\ell]$-valuation~or~a~single~valuation and have shown their 
approximation algorithms. 

Balcan, et al. \cite[Theorem 15]{BBCH-TR} consider the {\it tree\/} highway problem 
with a single valuation~as~a~natural~ex\-tension of the line highway problem 
with a single valuation and showed that it has~a~4-approximation~algo\-rithm. 
As a straightforward extension of Theorem 
\ref{thm-line-random}, we can easily show the following theorem for~the~tree 
highway problem with $[s,\ell]$-valuation. 
\begin{figure}[htb]
\fbox{
\begin{minipage}{17.25cm}
\begin{minipage}{17.0cm}\medskip
%
%
\makebox[1.3cm][r]{{\sf Input:}} A reduced instance $G=(V,E,\{w_{j}\})$ 
of the tree highway problem with $[s,\ell]$-valuation.\\
\makebox[1.3cm][r]{{\sf Output:}} A price vector 
$\bm{\sigma}=(\sigma_{1},\sigma_{2},\ldots, \sigma_{n})$ for $G$.
\begin{namelist}{~~2.}
\item[1.] Choose $r \in V$ arbitrarily as a root and 
construct the DAG representation 
$H=(U,F,\{w_{j}\})$ of $G=(V,E,\{w_{j}\})$. 
\item[2.] For each $u_{i} \in U$, assign a partial sum $s_{i} \in [0,\ell]$ 
for $u_{i}$ uniformly and at random. 
\item[3.] For each $i \in V$, compute a price $\sigma_{i}=s_{i}-s_{i-1}$ 
for the item $i$ 
and let $\bm{\sigma}=(\sigma_{1},\sigma_{2},\ldots, \sigma_{n})$. 
\end{namelist}
\end{minipage}\\[0.25cm]
\end{minipage}}
%
%
\caption{The algorithm {\sc Line\_Random}} \label{fig-tree-random}
\end{figure}
\begin{thm} \label{thm-tree-highway}
On an instance $G=(V,E,\{w_{j}\})$ of the tree highway problem 
with $[s,\ell]$-valuation,~the~algo\-rithm {\sc Tree\_Random} 
outputs a price vector $\bm{\sigma}$ that satisfies
\[
\msc{Opt}_{\rm coup}(G) \leq \frac{16}{3r}\cdot 
{\bf E}\left[\msc{Profit}_{\rm coup}(\bm{\sigma})\right]
\]
where $r=s/\ell$ is the ratio between the smallest and the largest valuations. 
\end{thm}

In this paper, we have focused on the highway problem under 
the coupon model,~however,~we~do not know much about the general item pricing 
problem under the other models such as the discount model, the 
$B$-bounded discount model, etc. 
So the interesting problem to be considered is 
\begin{namelist}{~~(3)}
\item[(1)] Design algorithms for the general item pricing problems 
under the ($B$-bounded) discount model. 
\end{namelist}
We are also interested in the inapproximability for the line and cycle highway 
problems with~$[s,\ell]$-valuation or a single valuation. Thus 
the interesting problem to be considered is 
\begin{namelist}{~~(3)}
\item[(2)] Derive the nontrivial lower bounds on the approximability for 
the line and cycle highway~problems with $[s,\ell]$-valuation or a single valuation. 
\end{namelist}
%

%
\end{document}